\documentclass[twocolumn,preprintnumbers,amsmath,amssymb,A4paper]{revtex4}

\usepackage{graphicx}
\usepackage{dcolumn}
\usepackage{bm}

\begin{document}
\newcommand{\kp}{{\bf k$\cdot$p}\ }
\title{The Klein paradox in a magnetic field: effects of electron spin}
\author{Wlodek Zawadzki and Pawel Pfeffer}
 \affiliation{Institute of Physics, Polish Academy of Sciences\\
 Al.Lotnikow 32/46, 02--668 Warsaw}

\begin{abstract}
Reflection and transmission of electrons scattered by a rectangular potential step in the presence of an external magnetic field parallel to the electron beam is described with the use of the Dirac equation. It is shown that in addition to the known effects present in the so called Klein paradox, the presence of magnetic field gives rise to  electron components with reversed spin in the reflected and transmitted beams. The spin-flip scattering processes are caused by the spin-orbit interaction activated by electric field of the potential step and transverse momentum components of electron motion induced by the magnetic field. The contemporary understanding of the Klein paradox, consisting in the finite transmission even when the potential height tends to infinity, is generalized to the presence of magnetic field and spin-reversed electron beams. The spin-reversed beams are shown to occur also for electrons moving above the step. It is proposed that, accounting for the anomalous value of the electron spin $g$-factor related to radiation corrections, the reflection and transmission scattering from the potential step can be used as an electron spin filter.
\end{abstract}

 \maketitle

\section{\label{sec:level1}Introduction\protect\\ \lowercase{}}

Exactly ninety years ago Klein [1] published his famous paper in which he described the reflection and transmission of relativistic electrons coming to a rectangular potential step using the relativistic equation which shortly before had been published by Dirac [2]. This paper is often referred to as the Klein paradox in which, seemingly, for a sufficiently high potential more electrons are reflected from the step than coming to it. However, Klein noted that Pauli had pointed out to him that if one takes into account the group velocity of the transmitted electron beam, one obtains $|R|^2 + |T|^2$ = 1, where $|R|^2$ and $|T|^2$ are the reflection and transmission probabilities, respectively, so there is no paradox. According to the contemporary views the paradox still exists and consists in the fact that, as the height of the step tends to infinity, the transmission coefficient tends to a non-zero limit. The paper of Klein was followed by important discussions and modified calculations, as reviewed in ref. [3].  Dombey and Calogeracos [4] concluded in their analysis that the theory of the Klein tunneling requires only the relativistic wave equations.

\begin{figure}
\includegraphics[width=8cm,height=8cm]{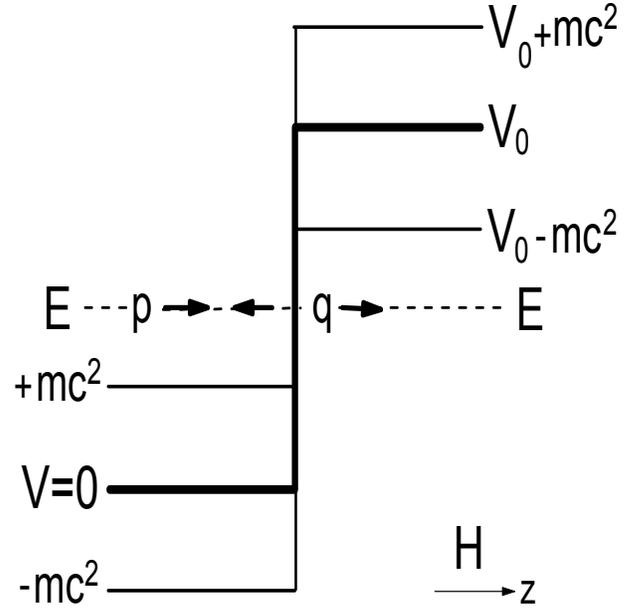}
\caption{Rectangular potential step illustrating the Klein scattering.
Electrons come from the left having the energy $E$ and momentum $p$ on the
left of the step and momentum $q$ on the right of the step. Magnetic field is
parallel to the $z$ direction. }
\label{fig1th}
\end{figure}

The standard Klein theory is limited to one dimension, parallel to the electron propagation (say, $z$). Our work treats the Klein scattering in the presence of an external magnetic field $\textbf{H}$ also parallel to $z$. The presence of the field induces the transverse motion components $p_x$ and $p_y$, so one has to do the theory in three dimensions.  The magnetic field naturally introduces to the problem the electron spin. The simultaneous presence of electric field (resulting from the potential step) and the external magnetic field activates the spin-orbit interaction inherently contained in the Dirac equation.  We concentrate on the spin effects caused  by $\textbf{H}$ and use the approach based on the Dirac equation. The sufficiency of our approach is confirmed by the theoretical work of Katsnelson et al [5] on the Klein tunneling in graphene, where the Dirac-type equations are sufficient to obtain important results.

The paper is organized as follows. Section II describes the theory.  Subsection \textbf{A} is concerned with the character of wave functions, \textbf{B} discusses a generalization of the Klein paradox, \textbf{C} treats the current conservation. Section III contains the discussion of results, the summary is given in the form of separate points. In the Appendix we enumerate and shortly discuss for completeness formulas for the standard case $H = 0$.

\section{\label{sec:level1} THEORY\protect\\ \lowercase{}}

We consider relativistic electrons coming from the left along the $z$ direction to the potential step described by the potential energy $V(z) = 0$ for $z < 0$ and $V(z) = V_0$ for $z  \ge 0$, in the presence of an external magnetic field also parallel to the $z$ direction. The investigated system is shown in Fig. 1. The stationary Dirac equation for the problem reads
\begin{equation}
\left[c \bm{\hat{\alpha}}\cdot \bm{\hat{P}}+mc^2\hat{\beta}) \right]\Psi(\textbf{r}) = (E-V)\Psi(\textbf{r})\;\;,
\end{equation}
where $\hat{\alpha} =(\alpha_x, \alpha_y, \alpha_z)$ and $\hat{\beta}$ are the standard Dirac $4 \times 4$ operators, $m$ is the rest electron mass,  $E$ is the energy and $\textbf{P} = \textbf{p} - e\textbf{A}/c$  is the kinetic momentum. The Dirac equation for electrons in a magnetic field alone has been solved by various authors. As in the absence of magnetic field, one deals with two spinors for positive electron energies and two for the negative ones. If the vector potential of magnetic field is taken in the asymmetric gauge \textbf{A} = $(-Hy, 0, 0)$, the eigenfunctions are given in terms of the harmonic oscillator functions $\Phi_n(\xi)$, where $\xi = (y -y_0)/L$ with $y_0 = k_xL^2$ and $L=(\hbar c/eH)^{1/2}$ being the magnetic radius. We use the explicit form of solutions following  Johnson and Lippmann [6].

The spin-up state (+,n) is
\ \\
\begin{equation}
\Theta(+,n) = A
\left(\begin{array}{c}(E-V+mc^2)\Phi_n\\0\\cp\Phi_n\\ (2mc^2 \cdot 2\mu_B H(n+1))^{1/2}\Phi_{n+1}\end{array}\right)
\;\;,
\end{equation}
with the energy
\ \\
$$
E(+,n) =\sqrt{c^2p^2+2mc^2\hbar\omega(n+1/2+1/2)+m^2c^4}+V =
$$
\begin{equation}
 = \sqrt{c^2p^2+C_{n+1}+m^2c^4}+V\;\;.
\end{equation}
\ \\
The spin-down state (-,n) is
 \ \\
\begin{equation}
\Theta(-,n) = A
\left(\begin{array}{c}0\\(E-V+mc^2)\Phi_{n}\\ (2mc^2\cdot 2\mu_BH n)^{1/2}\Phi_{n-1}\\-cp\Phi_{n}\end{array}\right)
\;\;,
\end{equation}
with the energy
\ \\
$$
E(-,n) = \sqrt{c^2p^2+2mc^2 \hbar\omega(n+1/2-1/2)+m^2c^4}+V =
$$
\begin{equation}
 = \sqrt{c^2p^2+C_n+m^2c^4}+V,
\end{equation}
\ \\
where $\omega = |e|H/mc$ is the cyclotron frequency  and  n = 0, 1, 2, ... . We use the notation $C_n=2mc^2 \hbar\omega n$ and $p_z = p$.

The energies  $E(+,n-1)$ and $E(-,n)$ are doubly degenerate with the exception of $E(-,0)$. This lowest energy is single and it does not depend on magnetic field:
\begin{equation}
E(-,0) = \sqrt{c^2p^2+m^2c^4}+V\;\;.
\end{equation}
\ \\
Here, as in Eqs. (3) and (5), $V = 0$ for $z < 0$ and $V = V_0$ for $z  \ge 0$.

At the left side of the step the momentum is
\begin{equation}
c^2p^2=E^2-(m^2c^4+C_n)\;\;.
\end{equation}
At the right of the step the momentum is
\begin{equation}
c^2q^2=(E-V_0)^2-(m^2c^4+C_n)\;\;.
\end{equation}

Suppose the electrons come to the barrier in the spin-up state (+,n-1). Since the reflection and transmission are elastic, there should exist two reflected waves (+,n-1) and (-,n) and two transmitted waves (+,n-1) and (-,n) having the same energy. Anticipating a little we  allow for the waves with the reversed spin. Thus we have, see Eqs. (2) and (4)
\ \\
\begin{widetext}
$$
\Psi(+,n-1) = \frac{1}{(2{\cal E}\cdot E)^{1/2}}\left[ e^{\frac{ip z}{\hbar}}
\left(\begin{array}{c}{\cal E}\Phi_{n-1}\\0\\cp\Phi_{n-1}\\C^{1/2}_{n}\Phi_{n}\end{array}\right)+R e^{\frac{-ip z}{\hbar}}
\left(\begin{array}{c}{\cal E}\Phi_{n-1}\\0\\-cp\Phi_{n-1}\\C^{1/2}_{n}\Phi_{n}\end{array}\right)+
R' e^{\frac{-ip z}{\hbar}}
\left(\begin{array}{c}0\\{\cal E}\Phi_{n}\\C^{1/2}_{n}\Phi_{n-1}\\cp\Phi_{n}\end{array}\right)
\right]_{z < 0}+
$$
\begin{equation}
\frac{1}{(2|{\overline{\cal E}}\cdot {\overline E}|)^{1/2}}\left[ Te^{\frac{iq z}{\hbar}}
\left(\begin{array}{c}{\overline{\cal E}}\Phi_{n-1}\\0\\cq\Phi_{n-1}\\C^{1/2}_{n}\Phi_{n}\end{array}\right)+T'e^{\frac{iq z}{\hbar}}
\left(\begin{array}{c}0\\{\overline{\cal E}}\Phi_{n}\\C^{1/2}_{n}\Phi_{n-1}\\-cq\Phi_{n}\end{array}\right)
\right]_{z\ge0}\;\;.
\end{equation}
\end{widetext}
\ \\
where we introduced the notation  ${\cal E}=E+mc^2$, ${\overline{\cal E}} = E+mc^2  - V_0$, ${\overline E} = E-V_0$.
The coefficients $R$ and  $R'$ are related to the reflected  waves with the same and reversed spins, respectively, while $T$ and $T'$ are those of the transmitted waves with the same and reversed spins, respectively. These coefficients can be determined by the boundary conditions, i.e. by equalizing each of the four spinor components in Eq. (9) at $z = 0$. This gives
\ \\
\begin{equation}
\frac{{\cal E}}{(2{\cal E}\cdot E)^{1/2}}(1+R)=\frac{\overline {\cal E}}{(2|{\overline{\cal E}}\cdot {\overline E}|)^{1/2}}T\;,
\end{equation}
\ \\
\begin{equation}
\frac{{\cal E}}{(2{\cal E}\cdot E)^{1/2}}R'=\frac{{\overline {\cal E}}}{(2|{\overline{\cal E}}\cdot {\overline E}|)^{1/2}}T'\;,
\end{equation}
\ \\
\begin{equation}
\frac{[cp(1-R)+C^{1/2}_{n}R']}{(2{\cal E}\cdot E)^{1/2}}=\frac{(cqT+C^{1/2}_{n}T')}{(2|{\overline{\cal E}}{\overline E}|)^{1/2}}\;,
\end{equation}
\ \\
\begin{equation}
\frac{[cpR'+C^{1/2}_{n}(1+R)]}{(2{\cal E}\cdot E)^{1/2}}=\frac{(C^{1/2}_{n}T-cqT')}{(2|{\overline {\cal E}}{\overline E}|)^{1/2}}\;,
\end{equation}

In order to simplify the subsequent formulas we introduce the so called kinematic factor $\kappa$. Employing Eqs. (7) and (8) one has
\begin{widetext}
\begin{equation}
\kappa = \frac{q{\cal E}}{p{\overline {\cal E}}} =
\frac{\{[E-V_0-(m^2c^4+C_n)^{1/2}][E-V_0+(m^2c^4+C_n)^{1/2}]\}^{1/2}(E+mc^2]}
{\{[E-(m^2c^4+C_n)^{1/2}][E+(m^2c^4+C_n)^{1/2}]\}^{1/2}(E-V_0+mc^2)}\;\;.
\end{equation}
\end{widetext}
\ \\
With the use of $\kappa$ we finally have
\ \\
\begin{equation}
R(+,n-1)=\frac{c^2p^2{\overline {\cal E}}^2(1-\kappa)(1+\kappa)-C_{n}V_0^2}{c^2p^2{\overline {\cal E}}^2(1+\kappa)^2+C_{n}V_0^2}
\end{equation}
\ \\
\begin{equation}
R'(+,n-1)=\frac{2cp{\overline {\cal E}}C^{1/2}_{n}V_0}{c^2p^2{\overline {\cal E}}^2(1+\kappa)^2+C_{n}V_0^2}
\end{equation}
\ \\
$$
T(+,n-1)=\frac{(|{\overline{\cal E}}\cdot {\overline E}|)^{1/2}{\cal E}}{({\cal E}\cdot E)^{1/2}{\overline{\cal E}}}[1+R(+,n-1)]=
$$
\ \\
\begin{equation}
= \frac{(|{\overline{\cal E}}\cdot {\overline E}|)^{1/2}}{({\cal E}\cdot E)^{1/2}}
\frac{2c^2p^2{\cal E}{\overline {\cal E}}(1+\kappa)}{{[c^2p^2\overline {\cal E}}^2(1+\kappa)^2+C_{n}V_0^2]}
\end{equation}
\ \\
$$
T'(+,n-1)=\frac{(|{\overline{\cal E}}\cdot {\overline E}|)^{1/2}}{({\cal E}\cdot E)^{1/2}}\frac{{\cal E}}{{\overline{\cal E}}}R'(+,n-1)=
$$
\ \\
\begin{equation}
=\frac{(\overline{\cal E}\cdot {\overline E})^{1/2}2cp{\cal E}C^{1/2}_{n}V_0}{({\cal E}\cdot E)^{1/2}[c^2p^2{\overline {\cal E}}^2(1+\kappa)^2+C_{n}V_0^2]}
\end{equation}
\ \\
where it is indicated that the coefficients are related to the initial (+,n-1) state.

It can be seen that in addition to the expected reflected  and transmitted components with the same spin, proportional  to $R$ and $T$, there exist the reflected and transmitted spin-reversed components proportional  to  $R'$ and $T'$. The latter are proportional to $C^{1/2}_{n}$, i.e. they vanish when there is no magnetic field. We indicate in the Discussion that the spin-flip components appear due to the spin-orbit interaction inherently present in the Dirac equation.

For the coming spin-down electrons in the state (-,n) the wave decomposition analogous to  Eq. (9) is
\begin{widetext}
$$
\Psi(-,n) = \frac{1}{(2{\cal E}\cdot E)^{1/2}}\left[ e^{\frac{ip z}{\hbar}}\left(\begin{array}{c}0\\{\cal E}\Phi_{n}\\C^{1/2}_{n}\Phi_{n-1}\\-cp\Phi_{n}\end{array}\right)+R e^{\frac{-ip z}{\hbar}}\left(\begin{array}{c}0\\{\cal E}\Phi_{n}\\C^{1/2}_{n}\Phi_{n-1}\\cp\Phi_{n}\end{array}\right)+
R' e^{\frac{-ip z}{\hbar}}\left(\begin{array}{c}{\cal E}\Phi_{n-1}\\0\\-cp\Phi_{n-1}\\C^{1/2}_{n}\Phi_{n}\end{array}\right)
\right]_{z < 0}+
$$
\begin{equation}
\frac{1}{(2|{\overline{\cal E}}\cdot {\overline E}|)^{1/2}}\left[ Te^{\frac{iq z}{\hbar}}\left(\begin{array}{c}0\\{\overline{\cal E}}\Phi_{n}\\C^{1/2}_{n}\Phi_{n-1}\\-cq\Phi_{n}\end{array}\right)
+T' e^{\frac{iq z}{\hbar}}\left(\begin{array}{c}{\overline{\cal E}}\Phi_{n-1}\\0\\cq\Phi_{n-1}\\C^{1/2}_{n}\Phi_{n}\end{array}\right)
\right]_{z\ge0}\;\;,
\end{equation}
\end{widetext}
\ \\
The coefficients are obtained again from the boundary  conditions. The results are the same as those given by Eqs.(15)-(18) with the reversed signs for $R'$ and $T'$.
\ \\
Thus also for the initial spin-down state (-,n) there exist after scattering the spin reversed components. However, for the lowest state (-,0) one obtains from Eqs. (16) and (18):  $R'(-,0) = 0$ and $T'(-,0) = 0$, i.e. the spin flips do not occur. This is understandable because for the corresponding lowest energy the electron has no transverse motion [see Eq. (6)] and in consequence the spin-orbit interaction is not activated.

\ \\
\subsection{CHARACTER OF WAVE FUNCTIONS}
\ \\

Now we consider the character of wave functions. It is related to relative values of the height  of the potential step $V_0$ and the electron energy $E$, see Fig. 1. It is clear that the components on the left of the step, i.e. in the $V = 0$ region, have the plain wave character. However, the components on the right of the step in the $V = V_0$ region can have either wave or decaying character. This depends  on the momentum $q$ in the $V_0$ region since $q$, given by Eq. (8), can be real or imaginary. In order to make the following discussion simpler and more transparent we neglect the magnetic field term $C_n$ in Eq. (8) as it is much smaller than $m^2c^4$. In this approximation $q$ is given by the relation
\ \\
\begin{equation}
c^2q^2 = [(E - V_0 - mc^2) (E - V_0 +mc^2)]^{1/2}\;\;.
\end{equation}

One can define three important cases:
\ \\

Case I: $V_0 - mc^2 >  E$ . Then $E+ mc^2 -V_0 < 0$ and $E - mc^2 -V_0 < 0$, so that $q$  is a real number. In consequence, the transmitted amplitudes in Eqs. (9) and (19) are plain waves. This means that electrons are partly reflected from the step, while the rest of them propagate from left to right inside the step.
\ \\

Case II. $V_0 + mc^2 < E$. Then $E - mc^2 - V_0 > 0$ and $E+ mc^2 - V_0 > 0$, so that $q$ is a real number. In consequence, the transmitted amplitudes in Eqs. (9) and (19) are plain waves. This means that electrons are partly reflected from the step, while in contrast to the case I, the rest of them propagate from left to right above the step.

Case III. $V_0 + mc^2 > E > V_0 - mc^2$. Then $E - V_0 - mc^2 < 0$ and $E - V_0 + mc^2 > 0$, so that $q$ is an imaginary number: $q = i|q|$ and the transmitted amplitudes are quickly decaying functions. This means that the electrons are completely reflected from the step while there are no electrons propagating inside the step.

The above considerations can be intuitively understood:  in the $V= V_0$ region the electrons cannot  propagate inside the gap $V_0 \pm mc^2$.
It is
clear that the above three cases can be introduced equally well without
neglecting the magnetic terms in Eq. (8).
Finally, we note that the quantity $\kappa$ being proportional to $q$, see
Eq. (14), has real  (cases I and II) or imaginary (case III) value.

\subsection{GENERALIZED KLEIN PARADOX}
\ \\

As already  mentioned, the contemporary understanding of the Klein paradox  for the step potential is that the transmitted wave  propagates even if the step's height tends to infinity. The probability densities of the transmitted waves are given by $|T|^2$  and $|T'|^2$. Let us consider the spin-conserved and spin-reversed transmission of the incoming electrons in the (+,n-1) state, as described by Eqs. (17) and (18), respectively. For the very high potential $V_0$ one can neglect smaller quantities in $|T|^2$  and $|T'|^2$ and in $\kappa$. The final results are
\ \\
\begin{equation}
|{T(+,n-1)}|^2  = \frac{(E+mc^2)4c^2p^2(cp+E+mc^2)^2}{E[(cp+E+mc^2)^2+C_{n}]^2}
\end{equation}
\ \\
and
\ \\
\begin{equation}
|{T'(+,n-1)}|^2 = \frac{(E+mc^2)4c^2p^2 C_{n}}{E[(cp+E+mc^2)^2+C_{n}]^2}
\end{equation}
\ \\
For the incoming electrons in (-,n) state the corresponding results for the very high $V_0$ limit are the same as those given above.

 It is seen that, even in the limit of infinitely high $V_0$, the probabilities $|T(+,n-1|^2$ and $|T'(+,n-1)|^2$ as well as $|T(-,n)|^2$ and $|T'(-,n)|^2$ do not vanish. The results of Eqs. (21) - (22) present  generalizations of the Klein paradox to the presence of magnetic field and spin-reversed electron beams.

\subsection{CURRENT CONSERVATION}

Now we check  the  conservation of the electron currents. We do this for the incident electrons in the state (+,n-1).
The sum of reflected and transmitted currents should be equal to the incident current.
\begin{equation}
j_{inc}=j_{R}+j_{R'}+j_{T}+j_{T'}
\end{equation}
where each current is given by the electron charge multiplied  by the probability density and the  group velocity $v_{gr} = \partial E/\partial p = c^2p/E$ for the currents in the $V = 0$ region and $v_{gr} = \partial {\overline E}/\partial q = c^2q/(E-V_0)$ in the $V = V_0$ region. Using Eq. (23) one has
\begin{equation}
\frac{ec^2p}{E}=\frac{e|R|^2c^2 p}{E}+\frac{e|R'|^2c^2 p}{E}+\frac{e|T|^2c^2 q}{(E-V_0)}+\frac{e|T'|^2c^2 q}{(E-V_0)}
\end{equation}

 With the use of Eqs. (17) and (18) we obtain
\begin{equation}
1=|R|^2+|R'|^2+\frac{{\cal E}{\overline E}|1+R|^2qE}{\overline {\cal E}Ep(E-V_0)}+\frac{{\cal E}{\overline E}|R'|^2qE}{\overline {\cal E}Ep(E-V_0)}
\end{equation}
which can be transformed into
\begin{equation}
1=|R|^2+|R'|^2+\kappa[|1+R|^2+|R'|^2]
\end{equation}
where $\kappa$ is given by Eq. (14). For the cases I and II, see subsection A, the quantity $\kappa$ is real and the first part of Eq. (26) is
\ \\
$$
|R|^2+|R'|^2 =
$$
\begin{equation}
=\frac{[c^4p^4{\overline {\cal E}}^4(1-\kappa^2)^2+C_{n}^2V_0^4+2c^2p^2{\overline {\cal E}}^2C_{n}V_0^2(1+\kappa^2)}  {[c^2p^2{\overline {\cal E}}^2(1+\kappa)^2+C_{n}V_0^2]^2}\;\;.
\end{equation}
\ \\
The second part is
\ \\
\begin{equation}
\kappa [|1+R|^2+|R'|^2]=\frac{4\kappa c^2p^2{\overline {\cal E}^2}C_{n}V_0^2+4\kappa c^4p^4{\overline {\cal E}}^4(1+\kappa)^2} {[c^2p^2{\overline {\cal E}}^2(1+\kappa)^2+C_{n}V_0^2]^2}\;\;.
\end{equation}
\ \\
 Adding the right-hand sides of Eqs. (27) and (28) and doing some algebra one obtains unity. The same result is obtained for the incoming current of electrons in the (-,n) state.

For the case III, $\kappa$ is imaginary: $\kappa=i|\kappa|$, the transmitted currents disappear and Eq. (26) takes the form
\begin{equation}
1=|R|^2+|R'|^2
\end{equation}
We check the above equality:
\ \\
$$
|R|^2+|R'|^2 =
$$
\begin{equation}
= \frac{[c^2p^2{\overline {\cal E}}^2(1+|\kappa|^2)-C_{n}V_0^2]^2+4c^2p^2C_{n}V_0^2}  {|[c^2p^2{\overline {\cal E}}^2(1-|\kappa|^2)+C_{n}V_0^2]+i[2c^2p^2{\overline {\cal E}}^2|\kappa|]|^2} = 1\;\;,
\end{equation}

This demonstrates the conservation of the currents in all three cases.

\section{\label{sec:level1} Discussion\protect\\ \lowercase{}}

As shown in Eqs. (9) and (19), the reflected and transmitted waves contain in addition to the standard spin-conserved contributions also the spin-reversed components. The reason for the appearance of spin flips is the spin-orbit interaction. It is well known that the relativistic Dirac equation contains inherently the spin-orbit interaction. This property is made explicit  by the Foldy-Wouthuysen expansion of the Dirac Hamiltonian up to the terms $v^2/c^2$, where $v$ is the electron velocity [7]. One than obtains the expression $H_{so}\sim (\bm{\hat{P}} \times \bm{\hat{\sigma}}) \cdot \partial V/\partial z$ in the standard notation, see ref. 8. In our case  the electric field $\partial V/\partial z$ is that of the potential step at $z$ = 0, while the transverse momentum components are provided by the magnetic field $\textbf{H} \parallel z$. This is easily seen since the  trajectories of electrons coming to the step for all the states higher than the lowest one (-,0) are spirals around the $z$ direction. Eqs. (16) and (18) show that the amplitudes of spin-flip components are directly related to the magnetic field via $C_n$ and they disappear for vanishing $H$. For the lowest state (-,0) the electron trajectory is not a spiral but a straight line, the transverse momenta vanish and the spin-flip wave components disappear, see the formulas after Eqs. (16) and (18).

There exists an interesting possibility allowing one to use the reflection from the potential step as an electron spin filter. It is known that the spin $g$-factor of an electron is not exactly 2 but 2.002319 due to radiation corrections  refs. [9, 10, 11]. This means that in the formulas (3) and (5) the spin terms are somewhat larger than the orbital ones, so that the (+,n-1) state is not degenerate with the (-,n) state but occurs somewhat higher. This difference in energy must be compensated by the lower $p_z$ value in the $p_z$ part of the energy [see Eq. (3)] since the total energy $E$ of the reflected electron must be the same as that of the incoming one. The different $p_z$ values are equivalent to different velocities $v_z = p_z/m(E)$, where $m(E)$ is the relativistic electron mass. In consequence, the reflected electrons with the reversed spins arrive to a given point at the left hand side of the step later than those with the conserved spins, so the two beams can be separated by an appropriate shut-down screens. The difference of arrival times for the two beams will depend on the distance between the step and the screen. The same reasoning can be made for the transmitted spin-conserved and spin-reversed electron waves.

\section{\label{sec:level1} Conclusions\protect\\ \lowercase{}}

We summarize our work in the form of conclusions.
\ \\

1. Scattering of relativistic electrons from a rectangular potential step in the presence of an external magnetic field is described with the use of the Dirac equation. The reflection and transmission amplitudes are calculated.

2. The electric field resulting from the presence of the step combined with the transverse components of electron motion caused by the magnetic field activate the spin-orbit interaction implicitly present in the Dirac equation. This interaction gives rise to the spin-flip reflection and transmission processes.

3. The spin-flip amplitudes in reflection and transmission  appear also for electron energies higher than the potential step.

4. The transmitted spin-conserved and spin-reversed electron beams have nonzero probability even for the step potential tending to infinity. This generalizes the Klein paradox to the presence of magnetic field and the spin-reversed currents.

5. It is shown in the general case that the sum of all reflected and transmitted electron currents is equal to the initial incoming current.

6. It is proposed that the anomalous electron spin $g$-value resulting from the radiation corrections can be used to employ the electron scattering from the step as a spin filter.

\appendix*
\section{}

Here we shortly enumerate and discuss the corresponding formulas for $H = 0$. It is done for completeness and  also because some of the formulas are difficult to find in the literature.
We put in all above formulas $C_n$ = 0. As already mentioned, for $H = 0$ the spin reversed currents do not appear since the spin-orbit interaction in the Dirac equation is not activated. In consequence there is $R' =  0,  T' = 0$. Now one has exactly the relation
\ \\
\begin{equation}
c^2q^2=(E-V_0-mc^2)(E-V_0+mc^2)\;\;.
\end{equation}
\ \\
i.e. identical with the approximate relation (20). The reasoning given in subsection A applies without changes, so that in cases I and II the quantities $q$ and $\kappa$ are real. This gives, see Eqs. (15) and (16),
\ \\
\begin{equation}
R=\frac{(1-\kappa)}{(1+\kappa)}\;\;\;\;\;\;\;T= \frac{(|{\overline{\cal E}}\cdot {\overline E}|)^{1/2}}{({\cal E}\cdot E)^{1/2}}
\frac{2{\cal E}}{{\overline {\cal E}}(1+\kappa)]}
\end{equation}

\ \\
In case III, $q$ and $\kappa$ are imaginary so that
\ \\
\begin{equation}
R=\frac{(1-i|\kappa|)}{(1+i|\kappa|)}\;\;\;\;\;T= \frac{(|{\overline{\cal E}}\cdot {\overline E}|)^{1/2}}{({\cal E}\cdot E)^{1/2}}
\frac{2{\cal E}}{{\overline {\cal E}}(1+i|\kappa|)]}
\end{equation}
\ \\
As to the current conservation,
since for $H=0$ the spin-flip currents do not exist, we have
\begin{equation}
j_{inc}=j_{R}+j_{T}
\end{equation}
which gives
\begin{equation}
\frac{ec^2p}{E}=\frac{e|R|^2c^2 p}{E}+\frac{e|T|^2c^2 q}{(E-V_0)}\;\;.
\end{equation}\ \\
Dividing both sides by $ec^2p/E$ we get
\begin{equation}
1=|R|^2+\kappa|1+R|^2
\end{equation}
\ \\
In cases I and II we have
\ \\
\begin{equation}
|R|^2+\kappa|1+R|^2=\frac{(1-\kappa)^2}  {(1+\kappa)^2}+\frac{\kappa(1+\kappa+1-\kappa)^2}{(1+\kappa)^2}=1\;\;,
\end{equation}
\ \\
while in case III Eq. (A.6) takes the form 1 = $|R|^2$, and we check, see Eq. (A.3)
\ \\
\begin{equation}
|R|^2 = |\frac{(1-i|\kappa|)}{(1+i|\kappa|)}|^2 = 1\;\;.
\end{equation}
\ \\
Thus for all the three cases the current conservation is fulfilled, see ref. 12.
\ \\

Now we consider the limit of $V_0\rightarrow\infty$, for which $q=-V_0/c$ and $\kappa=(-V_0)(E+mc^2)/[cp(-V_0)]=(E+mc^2)/cp$. This gives according to Eqs. (15) and (17)
\ \\
\begin{equation}
R=\frac{(E-mc^2)^{1/2}-(E+mc^2)^{1/2}}{((E-mc^2)^{1/2}+(E+mc^2)^{1/2}}
\end{equation}
\ \\
\begin{equation}
T= \frac{2cp}{E^{1/2}[(E+mc^2)^{1/2}+(E-mc^2)^{1/2}]}
\end{equation}
\ \\
i.e. the transmitted probability is
\ \\
\begin{equation}
|T|^2=\frac{2c^2p^2}{E(E+cp)}
\end{equation}
\ \\
which shows that for $V_0\rightarrow\infty$ the transmitted probability does not vanish. In contemporary understanding this is considered to be the Klein paradox.
\ \\
\ \\

\begin{thebibliography}
{99}\label{sec:TeXbooks}
\bibitem{pp1} O. Klein, Z. Phys.\textbf{ 53}, 157 (1929).
\bibitem{pp2} P. A. M. Dirac, Proc. Roy. Soc. \textbf{117}, 612 (1928).
\bibitem{pp3} A. Calogeracos and N. Dombey, Contemp. Phys. \textbf{40}, 313 (1999).
\bibitem{pp4} N. Dombey, A. Calogeracos, Physics Reports \textbf{315}, 41 (1999).
\bibitem{pp5} M. I. Katsnelson, K. S. Novoselov, and A. K. Geim, Nature
Phys. \textbf{2}, 620 (2006).
\bibitem{pp6} M. H. Johnson and B. A. Lippmann, Phys. Rev \textbf{76}. 828 (1949).
\bibitem{pp7} L. Foldy and S.A.  Wouthuysen, Phys. Rev. \textbf{78}, 29 (1950).
\bibitem{pp8} W. Zawadzki, Am. J. Phys. \textbf{73}, 756 (2005).
\bibitem{pp9} T. Kinoshita and M. Mio, Phys. Rev. Lett. \textbf{90}, 021803 (2003).
\bibitem{pp10} V.H. Hughes and T. Kinoshita, Rev. Mod. Phys. \textbf{71}, S133 (1999).
\bibitem{pp11} B. Odom, D. Hanneke, B. D'Urso and G. Gabrielse, Phys. Rev. Lett.\textbf{ 97}, 030801 (2006).
\bibitem{pp12} C. Itzykson and  J.B. Zuber, \emph{Quantum Field Theory }(McGraw-Hill, New York, 1985).
\end{thebibliography}
\end{document}